\documentclass[aps,pra,twocolumn,superscriptaddress,showpacs]{revtex4}
\def\be{\begin{equation}}
\def\ee{\end{equation}}
\def\bea{\begin{eqnarray}}          
\def\eea{\end{eqnarray}}
\def\bi{\begin{itemize}}
\def\ei{\end{itemize}}

\usepackage{graphicx}
\usepackage{xcolor}

\begin{document}

\title{ 
        Quench from Mott Insulator to Superfluid 
       }

\author{Jacek Dziarmaga} 
\affiliation{Instytut Fizyki Uniwersytetu Jagiello\'nskiego, 
             and Center for Complex Systems Research,
             ul. Reymonta 4, 30-059 Krak\'ow, Poland}

\author{Marek Tylutki}
\affiliation{Instytut Fizyki Uniwersytetu Jagiello\'nskiego, 
             and Center for Complex Systems Research,
             ul. Reymonta 4, 30-059 Krak\'ow, Poland}

\author{Wojciech H. Zurek}
\affiliation{Theory Division, Los Alamos National Laboratory, Los Alamos, NM 87545, USA,
             and
             Institut f\"ur Quantenphysik and Center for Integrated Quantum Science and Technology (IQST), 
             Universit\"at Ulm, Albert Einstein Allee 11, 89081 Ulm, Germany
             }

\date{ September 24, 2012 }

\begin{abstract}
We study a linear ramp of the nearest-neighbor tunneling rate in the Bose-Hubbard model driving the system from the Mott insulator state into the superfluid phase. We employ the truncated Wigner approximation to simulate linear quenches of a uniform system in 1,2, and 3 dimensions, and in a harmonic trap in 3 dimensions. In all these setups the excitation energy decays like one over third root of the quench time. The $-\frac13$ scaling arises from an impulse-adiabatic approximation - a variant of the Kibble-Zurek mechanism - describing a crossover from non-adiabatic to adiabatic evolution when the system begins to keep pace with the increasing tunneling rate.
\end{abstract}

\maketitle

\section{ Introduction }

Gapless quantum critical points are a serious obstacle for quantum simulation with ultracold atomic gases or ion traps, where one would like to prepare a simple ground state of a simple initial Hamiltonian and then drive the Hamiltonian adiabatically to an interesting final ground state. This general observation has been recently substantiated by a more quantitative theory \cite{QuantumKZ,review}, that is by now confirmed by several numerical studies \cite{sabbatini}. The theory is a quantum generalization of the classical Kibble-Zurek mechanism (KZM) \cite{K,Z}. The theory predicts that density of excitations (or excitation energy) decays with (usually a fractional) power of quench rate $1/\tau_Q$. Experiments dedicated to the quantum theory were made in Refs. \cite{ferro,Mercedes,deMarco1}. In Ref. \cite{deMarco1} ultracold spinless bosonic atoms in a three-dimensional (3D) optical lattice were driven from the Mott insulator phase to the superfluid phase across a quantum phase transition. The transition was non-adiabatic and the excitation energy was reported to decay with approximately the third root of the transition time. Much in the same spirit, adiabaticity of loading atoms into an optical lattice \cite{Kuba} or releasing them from the lattice confinement \cite{deMarco2} has been questioned recently. The 1/3-scaling reported in the experiment \cite{deMarco1} coincides with our earlier prediction in Ref. \cite{Meisner} for 1D chains. It is the aim of this paper to show that the same scaling also holds in 3D.

\begin{figure}[h!]
\includegraphics[width=1.15\columnwidth,clip=true]{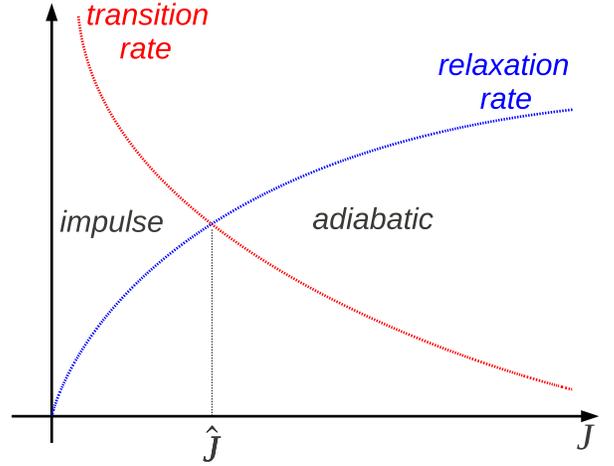}
\caption{ 
The transition rate $(dJ/dt)/J=J^{-1}\tau_Q^{-1}$ is huge at early time and small at late time. In contrast,
the relaxation rate $\tau^{-1}=J^{1/2}$ is negligible at early time and large at late time. The two rates are comparable
near $\hat J\simeq\tau_Q^{-2/3}$. Before $\hat J$ the evolution is impulse, i.e., the state does not change and remains the initial Mott state with random phases.
After $\hat J$ the evolution becomes adiabatic.
}
\label{FigImpulseAdiabatic}
\end{figure}

In this paper we generalize the theory developed in Ref. \cite{Meisner} for a 1D ring of BEC's to a two- and three-dimensional optical lattice in a harmonic trap potential. The system is initially prepared in the ground state deep in the Mott regime. Since atoms are localized in this regime, there is definite number of particles at each site that translates into an indefinite phase. The random phases at different sites are uncorrelated. After preparation of this state, the strength of the lattice potential is gradually reduced diminishing potential barriers between sites and increasing the nearest-neighbor tunneling rate. As the tunneling rate is ramped up towards the superfluid phase, the random initial phases become increasingly correlated. The range of these correlations as well as the excitation energy depend on the rate of this transition. We argue that the time evolution with the increasing tunneling rate $J$ can be roughly divided into two stages, see Fig. \ref{FigImpulseAdiabatic}. The first stage is non-adiabatic or impulse in the sense that the state of the system does not change and remains close to the uncorrelated initial state. After the instantaneous transition rate falls below the instantaneous relaxation rate of the system, at a $\hat J$, the evolution crosses over to the adiabatic stage. In this stage the phases become increasingly correlated. After the impulse-adiabatic crossover at $\hat J$ the system is in (local) thermal equilibrium. Both its temperature and excitation energy scale like the third root of the transition rate, apparently in agreement with the experiment in Ref. \cite{deMarco1}. However, in 2 and 3 dimensions even in the adiabatic stage the system has not enough time to develop the (quasi-)long-range order expected in thermal equilibrium at low temperature. This lack of long-range order manifests itself in the most spectacular way by topological vortex excitations. 

The paper is organized as follows. In Section \ref{BHmodel} we introduce the Bose-Hubbard model and define the linear quench (ramp) of the tunneling rate from the initial Mott state to the superfluid phase. In Section \ref{TWmethod} we briefly describe the truncated Wigner method \cite{TW} where bosonic operators are replaced by $c$-numbers, but expectation values are obtained as averages over stochastic realizations of the initial state. In Section \ref{Thermalization} we focus on the Josephson regime of relatively weak tunneling rate and use the truncated Wigner method to investigate thermalization of random initial states. We find that there is robust local thermalization in all relevant cases and estimate thermalization time. This is in agreement with previous studies of local thermalization after sudden change of tunneling parameter~\cite{eisert}. Motivated by this result, in Section \ref{Adiabatic} we consider an adiabatic process in the Josephson regime driven by a time-dependent tunneling rate and derive its corresponding adiabate equation. Given the estimate for thermalization time, in Sections \ref{IAC} and \ref{EE} we localize the crossover between the initial non-adiabatic (impulse) stage of the Mott-superfluid linear quench to the following adiabatic stage and estimate the initial temperature in the adiabatic process. The temperature as well as the excitation energy in the adiabatic stage are proportional to the third root of the quench rate. This power law is in agreement with the experiment \cite{deMarco1}. In Section \ref{Corr} we investigate one-particle correlation functions in the adiabatic stage. In 1D the correlation function is exponential as it should be in a thermal state. Its correlation length grows like the third root of the quench time. In contrast, in 2D and 3D we find that the function has thermalized at short distance, but it did not have enough time to develop the (quasi)-long-range order expected in thermal equilibrium at low temperature. 
This may be due to the topological defects left behind by the nonequilibrium transition at densities that are much higher than what would be expected from a thermal equilibrium at a corresponding final temperature.
Section \ref{Rabi} completes our discussion of the homogeneous case, i.e. without a harmonic trap, by extending the quench beyond the Josephson regime deep into the Rabi regime. Finally, in Section \ref{Trap} we repeat our simulations with a harmonic confinement and show that, as long as the initial Mott cloud extends over many lattice sites, the trap potential does not alter the third-root scaling of the excitation energy with the transition rate. We conclude in Section \ref{Concl}.

\section{ Bose-Hubbard model }\label{BHmodel}

The model describes spinless cold bosonic atoms in a $D$-dimensional optical cubic lattice \cite{Kasevich,Greiner,deMarco1,deMarco2} of $L^D$ sites numbered by a vector ${\bf s}\in {\cal Z}^D$. Its Hamiltonian reads
\bea
H = -J \sum_{\langle {\bf s_1},{\bf s_2} \rangle} a_{\bf s_1}^\dag a_{\bf s_2}
  + \sum_{\bf s}
      \left(  
              \frac{1}{2n}  a_{\bf s}^\dag a_{\bf s}^\dag a_{\bf s} a_{\bf s} + 
              V_{\bf s} a_{\bf s}^\dag a_{\bf s}
      \right)~.
\label{H}
\eea
Here $J$ is the hopping rate between nearest neighbor sites and $V_{\bf s}$ is a trap potential. For the uniform case, when $V_{\bf s}=0$, we assume periodic boundary conditions. In our units the interaction strength is $1/n$, where $n$ is the average number of atoms per site. In the thermodynamic limit there is a quantum phase transition from the Mott insulator to superfluid at $J_{cr}\simeq n^{-2}$. 

We drive the system by a linear quench 
\be
J(t)~=~{t}/{\tau_Q}~,
\label{Jt}
\ee
starting in the Mott ground state at $J=0$, 
\be
|n,n,n,\dots,n\rangle~,
\label{Mott}
\ee
with the same number of particles 
\be 
n\gg1
\ee 
at every site. We end either in the Josephson regime
\be
J~\ll~1~,
\ee
or extend the linear quench to the Rabi regime $J\gg1$.


\section{ Truncated Wigner method }\label{TWmethod}

When $n\gg1$ we can replace annihilation operators $a_{\bf s}$ by a complex field $\phi_{\bf s}$, 
$a_{\bf s}\approx\sqrt{n}\phi_{\bf s}$, normalized as $\sum_{\bf s} |\phi_{\bf s}|^2=L^D$ and evolving with the discrete Gross-Pitaevskii equation
\be
i\frac{d\phi_{\bf s}}{dt} = 
-J \nabla^2\phi_{\bf s} + 
\left(|\phi_{\bf s}|^2-1\right)\phi_{\bf s} ~,
\label{GPE}
\ee
see Refs. \cite{Amico,TW}. Here 
\be 
\nabla^2\phi_{\bf s}=\sum_{\alpha=1}^D\left(\phi_{\bf s+e_\alpha}-2\phi_{\bf s}+\phi_{\bf s-e_\alpha}\right)
\ee 
is a $D$-dimensional Laplacian. Truncated Wigner method was used to study dynamics of a quantum phase transition in Ref.~\cite{sabbatini}. For alternative approaches not using the truncated Wigner method see Refs. \cite{Kuba,sols}.

Quantum expectation values are estimated by averages over stochastic realizations of $\phi_{\bf s}$. Each realization has different random initial conditions coming from a Wigner distribution of the initial state (\ref{Mott}):
\be
\phi_{\bf s}(0)~=~e^{i\theta_{\bf s}(0)}~
\label{randomphases}
\ee
with independent random initial phases $\theta_{\bf s}(0)$. 

We consider kinetic and potential energy, 
\bea
E_{\rm kin} &=& J~\sum_{\bf s} \overline{ {\bf \nabla}\phi^*_{\bf s} {\bf \nabla}\phi_{\bf s} }~,\label{Ekin}\\
E_{\rm pot} &=& \sum_{\bf s} \frac12 \overline{\left(|\phi_{\bf s}|^2-1\right)^2} ~,\label{Epot}
\eea
where $\nabla_\alpha\phi_{\bf s}=\phi_{\bf s+e_\alpha}-\phi_{\bf s}$ and the overline means average over random initial conditions. 

\section{ Thermalization in Josephson regime }\label{Thermalization}

In this regime density fluctuations are relatively small, $|\phi_{\bf s}|^2\approx 1$, and it is convenient to parametrize 
\be 
\phi_{\bf s}~=~\left(1+f_{\bf s}\right)~e^{i\theta_{\bf s}}
\label{ftheta}
\ee
with real $f_{\bf s}$ and $\theta_{\bf s}$. After elimination of $f_s\ll 1$ in Eq. (\ref{GPE}) we obtain Josephson equations
\be
\frac{d^2\theta_{\bf s}}{dt^2}=
2J
\sum_{{\bf s'}}
\sin\left(\theta_{{\bf s}'}-\theta_{\bf s}\right)~,
\label{Josephson}
\ee
where the sum runs over sites ${\bf s'}$ that are nearest neighbors of ${\bf s}$. The initial conditions are random phases $\theta_{\bf s}(0)$ and vanishing velocities 
$\frac{d\theta_{\bf s}}{du}(0)=0$ equivalent to vanishing density fluctuations.

Since the parameter $J$ could be eliminated from Eq. (\ref{Josephson}) by introducing a rescaled time variable $u=J^{1/2}t$, the equations have a characteristic time-scale 
\be 
\tau ~\simeq~ J^{-1/2}~.  
\label{tau}
\ee
In particular, if there is relaxation towards thermal equilibrium, then $\tau$ is the thermalization time. 

\begin{figure}[h!]
\includegraphics[width=0.99\columnwidth,clip=true]{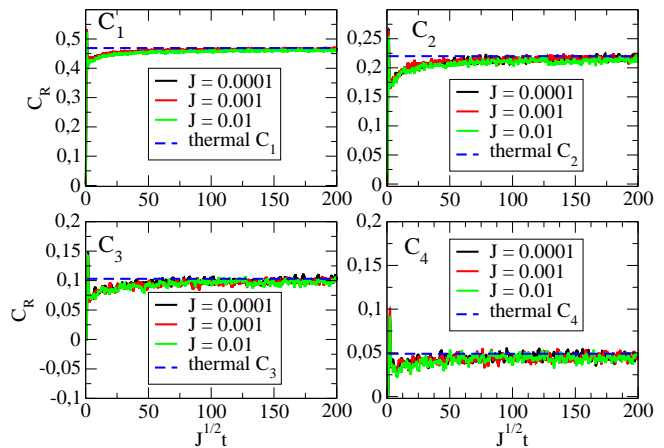}
\caption{ Thermalization in 1D ($L = 512$). In this figure we show correlation functions $C_R=\frac{1}{n}\langle a^\dag_s a_{s+R}\rangle=\overline{\phi^\star_s\phi_{s+R}}$ after an {\it instantaneous} quench from the initial Mott state (\ref{Mott},\ref{randomphases}) at $J=0$ to a final $J\ll1$ in the Josephson regime obtained with the truncated Wigner method. As predicted in Eqs. (\ref{Josephson},\ref{tau}), the three plots for the widely different final $J$ collapse in the rescaled time $J^{1/2}t$, proving that $\tau\simeq J^{-1/2}$ is indeed the characteristic time-scale in the Josephson regime. For large $J^{1/2}t$ the collapsed plots tend to their equilibrium values predicted in Eq. (\ref{A:CR}) in the Appendix, demonstrating that $\tau$ is indeed the thermalization time.
}
\label{FigRelaxation}
\end{figure}

\begin{figure}[h!]
\includegraphics[width=0.99\columnwidth,clip=true]{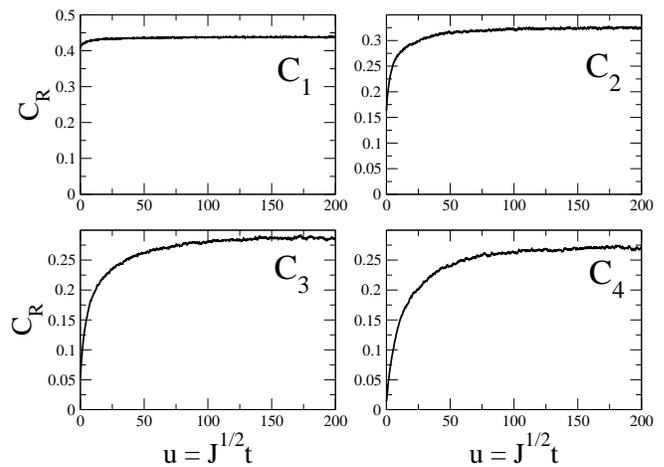}
\caption{ 
Thermalization in 3D ($128 \times 128 \times 128$ lattice). In this figure we show short-range correlators $C_1,\dots,C_4$ after an {\it instantaneous} quench from the initial Mott state (\ref{Mott},\ref{randomphases}) at $J=0$ to the final $J=0.1$ in the Josephson regime obtained with the truncated Wigner method. These short range correlators $C_R$ do thermalize but, unlike in 1D, their relaxation time clearly increases with $R$. There is quick local thermalization, but it takes longer time to thermalize the system over larger scales.
}
\label{C1234in3D}
\end{figure}

\begin{figure}[h!]
\includegraphics[width=0.9\columnwidth,clip=true]{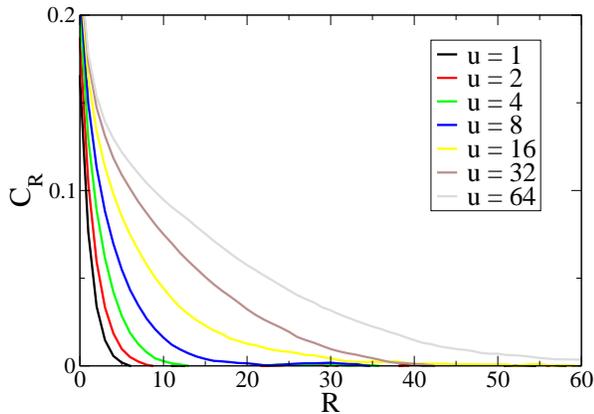}
\caption{ 
Thermalization in 3D. In this figure we show the correlation function $C_R(u)=\frac{1}{n}\langle a^\dag_s a_{s+R}\rangle=\overline{\phi^\star_s\phi_{s+R}}$ for several values of the rescaled time $u=J^{1/2}t$ at a fixed $J=0.01$. The initial state at $u=0$ has phase correlations of finite range. It is prepared by a linear quench from the initial Mott state (\ref{Mott},\ref{randomphases}) at $J=0$ to $J=0.01$ with $\tau_Q=512$. After this ramp, this low energy initial state thermalizes  at the fixed $J=0.01$ towards a low temperature thermal state with long-range order. While the short range correlations $C_R$ with small $R$ are quick to thermalize, see Fig. \ref{C1234in3D}, the long-range correlations are slow to develop the long range order expected at low temperature. As we can see in this figure, the range of $C_R$ roughly doubles when the time $u$ increases by a factor of $4$ i.e. the correlation range grows like the square root of time. 
}
\label{FigRelaxation3D}
\end{figure}

Thermalization in case of 1D is demonstrated in Fig. \ref{FigRelaxation}. A thermal state in 1D has finite correlation length at any finite temperature, hence the thermalization time is also finite. In contrast, in 3D at low temperature there is long-range order. These infinite-range correlations need infinite time to develop. Consequently, in 3D short range correlations are quick to thermalize, see Fig. \ref{C1234in3D}, but the range of long-range correlations grows roughly like the square root of time, see Fig. \ref{FigRelaxation3D}. However, the quick local equilibration is sufficient to thermalize local observables like the energy density.


In the Josephson regime $f_{\bf s}\ll 1$ in Eq. (\ref{ftheta}). Therefore the energy (\ref{Ekin},\ref{Epot}) is approximately quadratic
\bea
E_{\rm kin} &\approx & J~\sum_{\bf s} \nabla\theta_{\bf s}\nabla\theta_{\bf s} ~,\label{EkinJ}\\
E_{\rm pot} &\approx & \sum_{\bf s} 2f_{\bf s}^2  ~,\label{EpotJ}
\eea
At temperature $T$ the fields are distributed with $\exp[-(E_{\rm kin}+E_{\rm pot})/T]$ and the system is characterized by the averages 
\be 
\langle E_{\rm kin} \rangle ~=~ J\left(\frac{T}{2J}\right)~L^D,~~~
\langle E_{\rm pot} \rangle ~=~ \frac{T}{2}~L^D~
\label{EEJ}
\ee
satisfying the equipartition principle, and total energy 
\be 
\langle E\rangle=\langle E_{\rm kin} \rangle+\langle E_{\rm pot} \rangle=T L^D. \label{Etotal}
\ee

\section{ Adiabatic evolution }\label{Adiabatic}

When the process driven by the time-dependent $J$ in Eq. (\ref{Jt}) is adiabatic, then the system follows thermal equilibrium with a time-dependent $T$. On one hand, due to the changing temperature $T(t)$, its thermal energy $\langle E\rangle=TL^D$ changes at the rate
\be 
\frac{d}{dt}\langle E \rangle ~=~ \frac{dT}{dt} L^D~.
\label{dEdT}
\ee
On the other hand, the same energy changes due to the time-dependent Hamiltonian with the time-dependent $J$ at the rate
\be 
\frac{d}{dt}\langle E \rangle ~=~ \frac{dJ}{dt}~ \frac{\langle E_{\rm kin}\rangle}{J}~. 
\label{dEdJ}
\ee
Equating the two rates, (\ref{dEdT}) and (\ref{dEdJ}), we obtain a simple equation
\be 
\frac{dT}{dJ}~=~\frac{T}{2J}~.
\ee
Therefore $T=AJ^{1/2}$, with an integration constant $A$ depending on initial conditions, is the adiabate equation describing the adiabatic process.

\section{ Impulse-adiabatic crossover }\label{IAC}

In order to see when the evolution is adiabatic and when it is not, we must compare \cite{Z} the instantaneous transition rate
\be 
\frac{\frac{dJ}{dt}}{J}=\frac{1}{t}~=\frac{1}{J\tau_Q}~,
\ee 
which is huge at early time and small at late time, with the instantaneous relaxation rate in Eq. (\ref{tau})
\be 
\tau^{-1}\simeq J^{1/2}=\frac{t^{1/2}}{\tau_Q^{1/2}},
\ee
which is negligible at early time and large at late time. The two rates are comparable near
\be 
\hat J \simeq \tau_Q^{-2/3}
\label{hatJ}
\ee
at
\be 
\hat t \simeq \tau_Q^{1/3}~.
\label{hatt}
\ee
In a crude impulse-adiabatic approximation, see Fig. \ref{FigImpulseAdiabatic}, after $\hat J$ the evolution is adiabatic, but before $\hat J$ it is impulse in the sense that the state of the system does not change despite changing $J$ and the phases $\theta_{\vec s}$ remain as random as in the initial Mott state. Thus the random initial phases survive until the impulse-adiabatic crossover near $\hat J$, when they become initial conditions for the following adiabatic evolution. 

\section{ Excitation energy }\label{EE}

The impulse-adiabatic crossover takes place in the Josephson regime when $\hat J\ll 1$ or, equivalently, for slow enough quenches with
\be 
\tau_Q~\gg~J^{-3/2}~. \label{largetauQ}
\ee
Since at $\hat J$ the phases remain random, the kinetic energy in Eq. (\ref{EkinJ}) is 
$\langle E_{\rm kin} \rangle\simeq\frac{2\pi^2D}{3}\hat JL^D$. Comparing this with $\langle E_{\rm kin} \rangle=\frac12 TL^D$ in a thermal state, we obtain the initial temperature $T\simeq\frac{4\pi^2 D}{3}\hat J$ for the adiabatic process beginning at $\hat J$. This initial condition determines the constant $A$ in the adiabate equation $T=AJ^{1/2}$ as $A\simeq\frac{4\pi^2D}{3}\hat J^{1/2}$.
Consequently
\be 
T~\simeq~\hat J^{1/2}~J^{1/2}~\simeq~J^{1/2}\tau_Q^{-1/3}~ \label{TJosephson}
\ee
is the time-dependent temperature in the adiabatic process after $\hat J$. This solution remains accurate as long as $J\ll1$.

\begin{figure}[h!]
\includegraphics[width=1.0\columnwidth,clip=true]{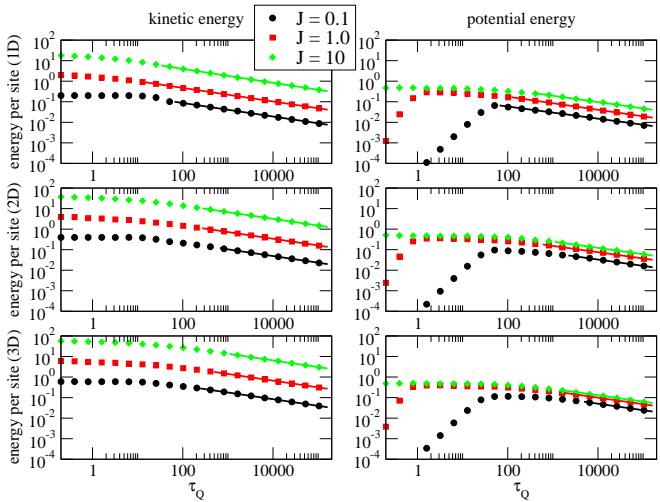}
\caption{
The figure shows dependence of the kinetic energy density (left column) and potential energy density (right column) for 1D (the upper row), 2D (the middle row) and 3D lattice (the bottom row). The lattice size $L=4096,256,128$ in $1D,2D,3D$ respectively. For large $\tau_Q\gg1$ we observe power law behavior consistent with the predicted $\langle E\rangle\sim \tau_Q^{-1/3}$. The best fits to the tails of the energy plots (the solid lines) give the exponents listed in Table \ref{TabKin}. 
The plots also demonstrate the equipartition $\langle E_{\rm kin} \rangle\approx\langle E_{\rm pot} \rangle$ at $J=0.1$ as predicted in the Josephson regime where the energy is approximately quadratic.
}
\label{FigNoTrap}
\end{figure}

In the adiabatic thermal state after $\hat J$ the kinetic and potential energies scale as
\be 
\langle E_{\rm kin} \rangle ~=~
\langle E_{\rm pot} \rangle ~=~ 
\frac12T L^D ~\simeq~ J^{1/2}\tau_Q^{-1/3}L^D ~, \label{ET}
\ee
in consistency with the numerical data in Figure~\ref{FigNoTrap} and Table \ref{TabKin}. 

\begin{table}[h!]
\begin{tabular}{|c||c|c|c||c|c|c||}
\hline
$J$ & 
$\langle E_{\rm kin}^{\rm 1D}\rangle$ & 
$\langle E_{\rm kin}^{\rm 2D}\rangle$ & 
$\langle E_{\rm kin}^{\rm 3D}\rangle$ & 
$\langle E_{\rm pot}^{\rm 1D}\rangle$ & 
$\langle E_{\rm pot}^{\rm 2D}\rangle$ & 
$\langle E_{\rm pot}^{\rm 3D}\rangle$ 
\\ \hline \hline
$0.1$  & $0.33$ & $0.33$ & $0.33$ & $0.33$ & $0.30$ & $0.30$\\
$1.0$  & $0.33$ & $0.33$ & $0.33$ & $0.33$ & $0.31$ & $0.31$\\
$10.0$ & $0.33$ & $0.33$ & $0.33$ & $0.32$ & $0.31$ & $0.31$\\
\hline
\end{tabular}
\caption{ 
The best fits to $\alpha$ in $\langle E_{\rm kin/pot}\rangle\sim\tau_Q^{-\alpha}$ are consistent with $\alpha=1/3$. Lattice sizes as in Fig. \ref{FigNoTrap}. 
}
\label{TabKin}
\end{table}

\section{ Correlations }\label{Corr}

\begin{figure}[h!]
\includegraphics[width=0.99\columnwidth,clip=true]{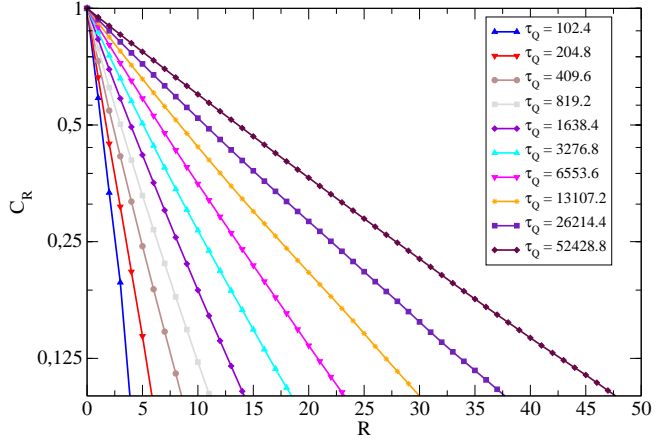}
\caption{ 
Correlation functions $C_R$ in 1D at $J=0.1$ for different quench times $\tau_Q$ and the lattice length $L = 4096$. The functions are exponential, as expected in a thermal state in the adiabatic stage of the evolution. Their correlation length scales like $\xi\sim\tau_Q^\alpha$ with the best fit $\alpha=0.329$ close to the predicted $1/3$.
}
\label{FigCR1D}
\end{figure}

In 1D a thermal correlation function is exponential, see Eq. (\ref{CR1D}) in the Appendix,
\be 
C_R ~=~
\frac{1}{n}
\langle a_{\vec s}^\dag a_{\vec s+\vec R} \rangle ~=~
\exp\left(-R/\xi\right)
\label{CR}
\ee
with a time-dependent correlation length
\be 
\xi ~\approx~ \frac{4J}{T} ~\simeq~ \frac{J^{1/2}}{\hat J^{1/2}} ~\simeq~ J^{1/2} \tau_Q^{1/3}~,
\label{xi}
\ee
compare Eq. (\ref{xiT}). In the above derivation of $\xi$ we used Eqs. (\ref{TJosephson}, \ref{ET}), which are valid for thermal equilibrium in the Josephson regime, i. e. when $J \ll 1$. 
At $\hat J$, when the phases are still as uncorrelated as in the initial Mott state, the length is comparable to the lattice constant $1$, but after $\hat J$ it grows like $J^{1/2}$: the system is ordering by increasing the range of correlations. At any given $J$ in the range $\hat J\ll J\ll1$ the correlation length scales like $\tau_Q^{1/3}$, i.e., the system is correlated more for slower quenches. This prediction is consistent with the numerical data in Fig. \ref{FigCR1D}. 

\begin{figure}[h!]
\includegraphics[width=0.99\columnwidth,clip=true]{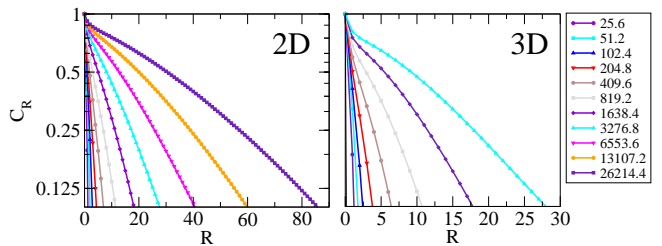}
\caption{ 
Long-range tail of the correlation functions $C_R$ in 2D (lattice size $L = 256$) and 3D (lattice size $L = 128$) at $J=0.1$ for different quench times $\tau_Q$. They are in the adiabatic stage, in the sense that local observables have equilibrated, but they have not had enough time to develop 
infinite-range (quasi-)long-range order expected in the low temperature phase. Instead the correlations have finite range limited by a finite rate at which they can spread across the system. The range grows with $\tau_Q$ faster than $\tau_Q^{1/2}$. 
}
\label{FigCRin23D}
\end{figure}

In 1D in thermal equilibrium there is finite correlation length $\xi$ and, consequently, finite relaxation time. The system can reach thermal equilibrium in finite time because it needs to order only up to the finite distance $\xi$. 
In contrast, in 2D and 3D the correlation function $C_R$ in low temperature thermal equilibrium either decays with a power of the distance $R$ (quasi-long-range order in 2D) or tends to a constant (long-range order in 3D). In either case the equilibrium correlations have infinite range. For a system initialized with random phases it is impossible to build up such infinite-range correlations in a finite time proportional to $\tau_Q$. Thus the system does not reach thermal equilibrium at all length scales: it is correlated as in a thermal state up to a finite range, but it remains uncorrelated at longer distances. The short range thermal correlations explain the $-\frac13$ scaling of the excitation energy, because the energy is a local observable not sensitive to the long range correlations. This is the main result of our paper that is experimentally relevant and strongly supported by the numerics summarized in Table I.

The long range correlations in 2D/3D are presented in Fig. \ref{FigCRin23D}. These correlation functions show how the system is equilibrating in time. We do not have analytic predictions for these non-equilibrium functions. Apparently they are not exponential, so it would be pointless to fit them with an exponent to find correlation lengths, but one can roughly estimate that their range defined as, say, the $R$ where $C_R$ falls below $0.25$, grows faster than $\tau_Q^{1/2}$.  

\begin{figure}[!h]
\includegraphics[width=0.68\columnwidth,clip=true]{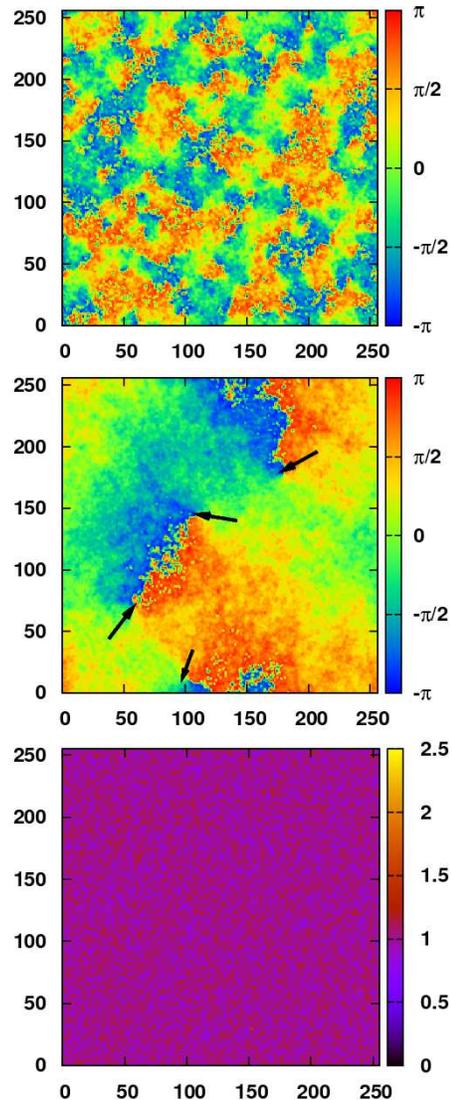}
\caption{
Quench on a 2D periodic $256\times256$ lattice in the Josephson regime at $J=0.1$.
Top panel: phase $\theta_{\bf s}$ for a quench time $\tau_Q=1638.4$.
Middle panel: phase $\theta_{\bf s}$ for a quench time $\tau_Q=52428.8$.
Bottom panel: atomic density $|\phi_{\bf s}|^2$ associated with the phase in the middle panel.
The top and middle panel demonstrate that the range of phase correlations increases 
with increasing $\tau_Q$. Average size of domains of constant phase is consistent with the
range of 2D correlations in Fig. \ref{FigCRin23D}. 
In the faster quench (top panel), where the range of correlations is comparable to the lattice spacing, there is plenty of random vorticity. 
In the slower quench (middle panel), where the range of correlations is much longer than the lattice spacing, it is possible to identify smooth topological vortices (marked by the arrows). 
The associated density fluctuations in the bottom panel are small, within $10\%$ of the average density $|\phi_{\bf s}|^2=1$, and very weakly correlated between different sites. 
This is not really surprising because in the Josephson regime the healing length $\simeq J^{1/2}$ is less than the lattice spacing and, consequently, densities at different sites are only weakly coupled and vortex cores are less than the lattice spacing. 
}
\label{Vortex2D}
\end{figure}

\begin{figure}[!h]
\includegraphics[width=0.8\columnwidth,clip=true]{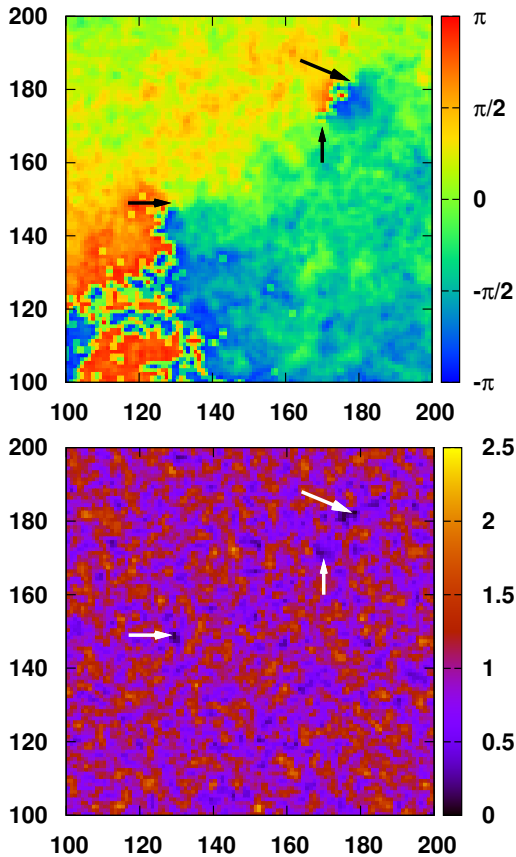}
\caption{
A quench with $\tau_Q=3276.8$ at $J=1$ on a part of a 2D periodic $256\times256$ lattice.
In the top panel the phase $\theta_{\bf s}$ and in the bottom panel the density distribution
$|\phi_{\bf s}|^2$. The three arrows mark an isolated vortex and a vortex-antivortex pair. 
Unlike in the Josephson regime at $J=0.1$ where density fluctuations are small and vortex cores 
are less than the lattice spacing, see Fig. \ref{Vortex2D}, here at $J=1$ the topological vortices 
in the upper panel are associated with clear dips in atomic density marked in the bottom panel. 
}
\label{Vortex2DJ1}
\end{figure}

What is more interesting, the absence of the long-range order or, equivalently, finite correlation range leaves open the possibility of topological vortex excitations, see Figures~\ref{Vortex2D} and \ref{Vortex2DJ1} at $J=0.1$ and $J=1$ respectively. In the Josephson regime at $J=0.1$ the healing length $\simeq J^{1/2}$ is less than the lattice spacing $1$. Consequently, there are only small and weakly correlated density fluctuations around the average density $|\phi_{\bf s}|^2=1$ and there are no dips in atomic density associated with the topological vortices in Fig.~\ref{Vortex2D}, i.e., their cores are thinner than the lattice spacing. In contrast, at $J=1$ when the healing length becomes longer than the lattice spacing topological vortices develop empty cores, compare the upper and bottom panels in Fig.~\ref{Vortex2DJ1}. The isolated topological vortices in 2D shown in Figs. \ref{Vortex2D} and \ref{Vortex2DJ1} are excitations above the thermal equilibrium in the Berezinski-Kosterlitz-Thouless phase \cite{Sodano}. A different perspective on vortex formation is provided by Ref.~\cite{sola}, where vorticity in both two and three dimensional XY model is discussed.


\section{ Into the Rabi regime }\label{Rabi}

The linear quench can be extended beyond the Josephson regime into the Rabi regime. When 
\be 
J~\gg~1
\ee 
the hopping term dominates over the nonlinear interaction in Eq.~(\ref{GPE}), but the nonlinearity is essential to keep the system thermalized. In the Rabi regime
$\langle E \rangle\approx\langle E_{\rm kin} \rangle$ Eq. (\ref{dEdJ}) becomes 
\be 
\frac{d}{dt}\langle E_{\rm kin} \rangle =  \frac{dJ}{dt} \frac{\langle E_{\rm kin} \rangle}{J}~.
\ee  
Consequently $\langle E_{\rm kin} \rangle\propto J$. The proportionality factor is fixed by the initial kinetic energy right after the Josephson/Rabi crossover near $J\simeq 1$. This energy is roughly equal to the final energy when the system is leaving the Josephson regime: $\langle E \rangle\simeq\tau_Q^{-1/3}L^D$, see Eq. (\ref{ET}) with $J\simeq1$.
Thus the dominant kinetic energy in the Rabi regime is
\be 
\langle E \rangle~\approx~\langle E_{\rm kin}\rangle~\simeq~J\tau_Q^{-1/3}L^D~. \label{ERabi}
\ee
It is linear in the time-dependent $J$ and scales like $\tau_Q^{-1/3}$, compare Fig. \ref{FigNoTrap}. The scaling (\ref{ERabi}) is valid provided that the impulse-adiabatic crossover takes place in the Josephson regime: $\hat J\ll1$ or, equivalently, $\tau_Q\gg1$.

For faster quenches with $\tau_Q\ll1$ the impulse stage extends into the Rabi regime, where the thermalization rate $\tau^{-1}\simeq1$ is set by the strength of the nonlinearity in Eq. (\ref{GPE}). It becomes comparable to the transition rate $J^{-1}dJ/dt=1/t$ at $\hat t\simeq1$ or, equivalently, $\hat J\simeq\tau_Q^{-1}\gg1$ when the impulse stage terminates. 
At $\hat J$ the phases remain as random as in the initial state and the kinetic energy is 
$\left.\langle E_{\rm kin} \rangle\right|_{\hat J}\simeq\hat JL^D$. In the following adiabatic stage, this dominant kinetic energy scales with increasing $J$ like 
\be 
\langle E_{\rm kin}\rangle~\simeq~
\frac{J}{\hat J}~
\left.\langle E_{\rm kin} \rangle\right|_{\hat J}~\simeq~
JL^D~,
\ee
i.e., the excitation energy does not depend on the quench time $\tau_Q$ when $\tau_Q\ll1$. This is consistent with the numerical data in Fig. \ref{FigNoTrap}.

\section{ Linear quench in a 3D harmonic trap }\label{Trap}

\begin{figure}[h!]
\includegraphics[width=0.8\columnwidth,clip=true]{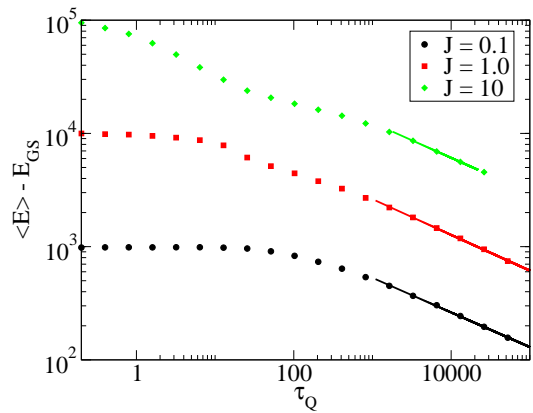}
\caption{
Excitation energy above the discrete Gross-Pitaevskii ground state at a given $J$, $\langle E \rangle - E_{\rm GS}(J)$, as a function of $\tau_Q$ for a 3D lattice in a harmonic trap. The tails $\langle E \rangle - E_{\rm GS}\sim\tau_Q^{-\alpha}$ for large $\tau_Q\gg1$ are best fitted with the exponents: $\alpha=0.31$ for $J = 0.1$, $\alpha=0.32$ for $J = 1.0$, and $\alpha=0.31$ for $J = 10.0$. Here the lattice size is $64^3$ and the initial Thomas-Fermi radius $R_{\rm TF}=10$. 
}
\label{Figtrap}
\end{figure}

\begin{figure}[h!]
\includegraphics[width=0.99\columnwidth,clip=true]{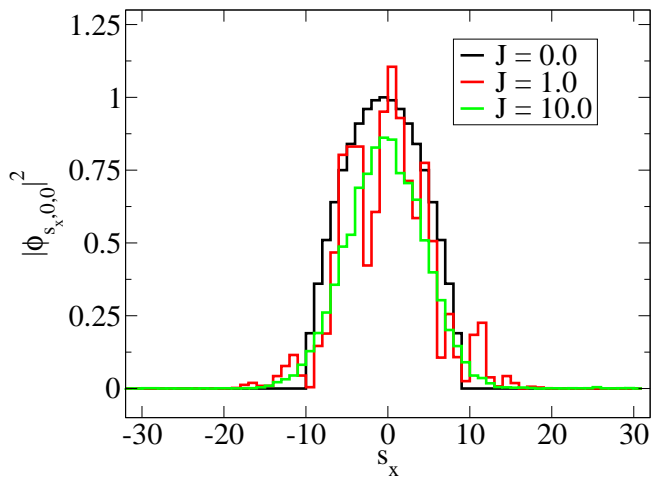}
\caption{ 
A cross-section of $|\phi_{\bf s}|^2$ along the line $s_y=s_z=0$ in a single realization of the statistical ensemble for $\tau_Q=10$. The initial Thomas-Fermi profile at $J=0$ spreads into a near-Gaussian wave-packet at $J=10$. The simulation performed for a 3D lattice of $64 \times 64 \times 64$ sites. 
}
\label{FigCloud}
\end{figure}

\begin{figure}[h!]
\includegraphics[width=0.86\columnwidth,clip=true]{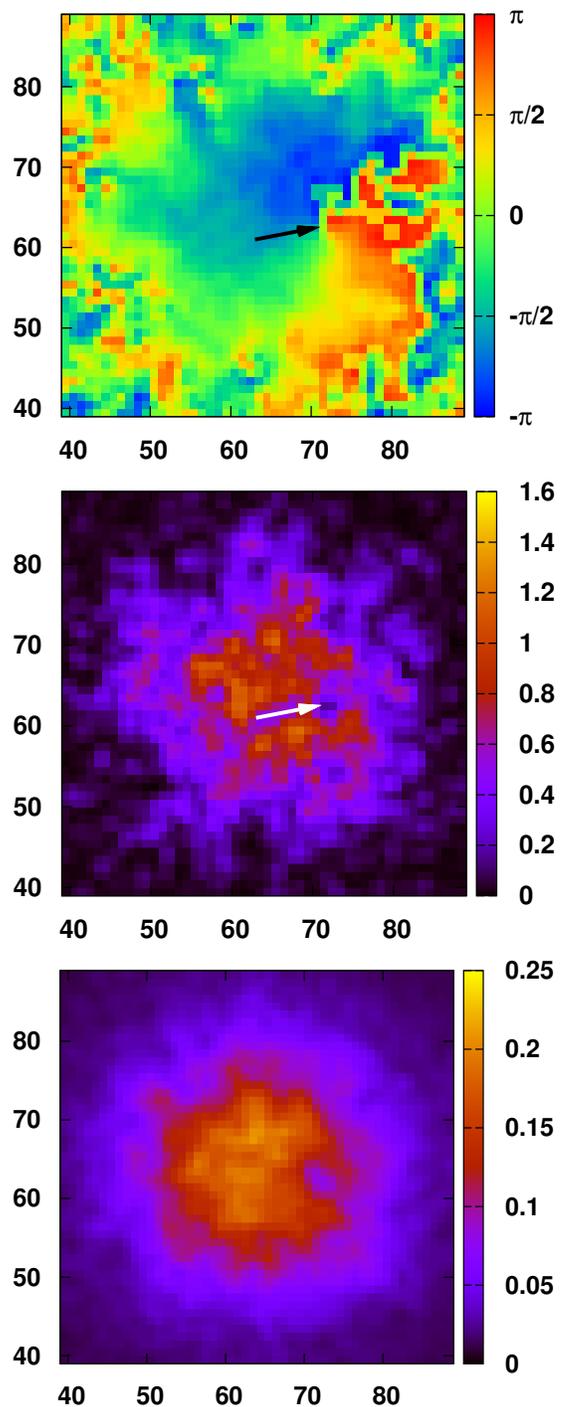}
\caption{
Linear quench in the 3D trap with $\tau_Q=204.8$ and the initial Thomas-Fermi radius $R_{\rm TF}=25$
at the tunneling rate $J=1$ for $128 \times 128 \times 128$ lattice.
Top panel: phase in the $x-y$ cross-section across the center of the trap.
Middle panel: density in the $x-y$ cross-section across the center of the trap.
Bottom panel: cumulative column density in the direction perpendicular to the $x-y$ plane.
The arrows in the top and middle panel point to a vortex.
}
\label{Vortex3DColumn}
\end{figure}

In a harmonic trap the discrete Gross-Pitaevskii equation (\ref{GPE}) becomes
\be
i\frac{d\phi_{\bf s}}{dt} = 
-J \nabla^2\phi_{\bf s} + 
\frac12 \omega^2 {\bf s}^2 \phi_{\bf s} +
|\phi_{\bf s}|^2\phi_{\bf s}~.
\label{GPEtrap}
\ee
The initial state at $J=0$ has random phases $\theta_{\bf s}(0)$, as in the uniform case, and a Thomas-Fermi density profile
\be 
|\phi_{\bf s}(0)|^2 = \frac{\omega^2}{2} \left( R^2_{\rm TF} - {\bf s}^2 \right)
\label{TF}
\ee
for sites ${\bf s}$ inside a sphere of radius $R_{\rm TF}$ and zero otherwise. In order to make comparisons easier, we set $\omega^2=\frac{2}{R_{\rm TF}^2}$ here to have $|\phi_{\bf s}(0)|^2=1$ in the center of the trap just as in our uniform calculations. Numerical results, collected in Figs. \ref{Figtrap} and \ref{FigCloud}, demonstrate that in the trap the excitation energy also scales with an exponent close to $-\frac13$ just as in the uniform case. At the larger $J=1$ there are large density
fluctuations, see Figs. \ref{FigCloud} and \ref{Vortex3DColumn}, and plenty of random vorticity, see Fig. \ref{Vortex3DColumn}.


\begin{figure}[h!]
\includegraphics[width=0.8\columnwidth,clip=true]{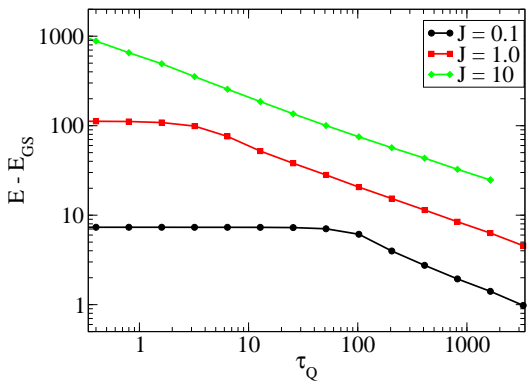}
\caption{
The same excitation energy as in Fig. \ref{Figtrap} but with constant initial phases $\theta_{\bf s}=0$. For large $\tau_Q$ the energy decays approximately like 
$\tau_Q^{-0.5}$ at $J=0.1$ and $\tau_Q^{-0.4}$ at $J=10$ i.e. faster than for the random initial phases in Fig. \ref{Figtrap}. The size of the lattice is $64 \times 64 \times 64$. 
}
\label{FigConstPhase}
\end{figure}

The key difference with respect to the uniform case is that turning on the tunneling rate $J$ makes the trapped cloud expand with respect to the original Thomas-Fermi profile \cite{Haque}, see Fig. \ref{FigCloud}. In an attempt to isolate the effect of expansion from the (uniform) Kibble-Zurek mechanism, we rerun our simulations with a constant initial phase $\theta_{\bf s}(0)=0$ across the system instead of the usual random initial phases (\ref{randomphases}). The resulting excitation energies are shown in Fig. \ref{FigConstPhase}. They are not only a factor of $10^2$ lower than in the corresponding Fig. \ref{Figtrap}, but also their decay with $\tau_Q$ is steeper, i.e., with an exponent closer to $-1/2$ than to the ``random'' exponent $-\frac13$. This is not too surprising, as the kinetic energy density of the random phases, $\simeq J$, far outweights average density of kinetic energy in the Thomas-Fermi profile with a constant phase, $\simeq JR^{-2}_{\rm TF}$, for any reasonable $R_{\rm TF}\gg1$.

The exponent $-1/2$ for the constant initial phase can be explained by an impulse-adiabatic argument again. When $J\ll1$ the Thomas-Fermi profile (\ref{TF}) is a good approximation to the ground state of the discrete Gross-Pitaevskii equation. When 
$R_{\rm TF}\gg1$ its lowest Bogoliubov excitation is
\be 
\delta\phi_{\bf s}(t)~=~
a~i~
\sin\left(\omega_1t+\varphi\right)~
j_0\left(\frac{\pi|{\bf s}|}{R_{\rm TF}}\right)
\ee
with real amplitude $a$ and frequency $\omega_1=J\pi^2R^{-2}_{\rm TF}$. Here $j_0(x)$ is the spherical Bessel function. It is a breathing mode describing radial flows of particles. In a linear quench of the tunneling rate, $J(t)=t/\tau_Q$, the evolution is impulse as long as the transition rate $\frac{dJ/dt}{J}=1/t$ is much less than the $\omega_1$, i.e., up to 
$\hat J\simeq R_{\rm TF}\tau_Q^{-1/2}$. At $\hat J$ the wavefunction is still the initial Thomas-Fermi profile (\ref{TF}) with a constant phase, but the profile is no longer the ground state of the discrete Gross-Pitaevskii equation and its excitation energy with respect to the ground state is 
\be 
E-E_{\rm GS} ~\simeq~ \hat J R_{\rm TF} ~\simeq~ R^2_{\rm TF} \tau_Q^{-1/2}.
\ee 
It decays with $\tau_Q$ with the exponent $-1/2$. It is also much less than the corresponding excitation energy at the $\hat J$ for random initial phases, $\langle E\rangle-E_{\rm GS}\simeq \hat J R^3_{\rm TF}$, for any reasonable $R_{\rm TF}\gg1$.  

\section{ Conclusion }\label{Concl}

Our results justify the following simple picture. A linear ramp of the tunneling rate at first takes the system by surprise: the ramp is too fast, or the system too slow, for the initial Mott state to adjust to the increasing tunneling rate. In this impulse stage phases at different lattice sites remain as uncorrelated as in the initial Mott state, but the frozen Mott state gradually deviates from the instantaneous ground state. At some point, however, reactions of the system become fast enough to catch up with the ramp and the non-integrable system thermalizes locally. In the following adiabatic process, its excitation energy (or temperature) scales like $\tau_Q^{-1/3}$. This mechanism is quite insensitive to the trapping potential because the kinetic energy accumulated in the initial random phases typically far exceeds the kinetic energy due to localization by harmonic confinement. The absence of thermalization at large scale manifests itself by topological vortex excitations.

{\bf Acknowledgments. ---} 
This work was supported in part by 
the NCN grant DEC-2011/01/B/ST3/00512 (JD,MT),
the DoE via LDRD program at the Los Alamos National Laboratory (WHZ), 
and the PL-Grid Infrastructure (MT).

\section*{Appendix: Thermalization in the Josephson regime after a sudden quench in 1D}

The Josephson equations (\ref{Josephson}) follow from a Hamiltonian
\be 
H_J=
\sum_{\bf s}\frac{p_{\bf s}^2}{2}+
\sum_{\bf s}\sum_{\alpha=1}^D
2J
\left[
1-\cos\left(\theta_{\bf s+e_\alpha}-\theta_{\bf s}\right),
\right]
\ee
where $p_{\bf s}=\dot\theta_{\bf s}$. A thermal state is given by a factorizable Boltzmann probability distribution $f(p,\theta)=f_p(p)f_\theta(\theta)\propto\exp(-H_J/T)$. 
Here $f_p$ is a Gaussian and, in case of a 1D chain, 
\bea 
f_\theta(\theta) &\propto& 
\exp\left[2J\sum_s\cos\left(\theta_{s+1}-\theta_s\right)/T\right]
\nonumber\\
&\equiv&
\prod_s
\exp\left[2J\cos\Delta\theta_s/T\right]~.
\eea
For a chain much longer than a correlation length, $f_\theta$ can be approximately factorized into a product of distributions for independent random phase steps $\Delta\theta_s=\theta_{s+1}-\theta_s$. 

The product can be used e.g. to calculate thermal correlation functions
\be 
C_R=
\overline{e^{i\theta_{s+R}}e^{-i\theta_s}}=
\prod_{s=0}^{R-1} \overline{e^{i\Delta\theta_s}}=
\left[\frac{I_1(2J/T)}{I_0(2J/T)}\right]^R~,
\label{CR1D}
\ee
where $I_m$ is the modified Bessel function. The correlation length is
\be 
\xi=1/\log\left[\frac{I_0(2J/T)}{I_1(2J/T)}\right]~\approx~\frac{4J}{T}~
\label{xiT}
\ee 
with the last approximation for small $T\ll4J$. 

We are in a position now to analyze the instantaneous quench in Figure \ref{FigRelaxation} from $J=0$ to a finite $J\ll1$. The Mott state is the initial state right after the quench.
It is characterized by $p_s(0)=0$ and random $\theta_s(0)$ and, consequently, its average energy per site is $E_{\rm in}/L=2J$. This energy is conserved in the following evolution with Josephson equations as the system thermalizes to a temperature $T$ with average energy per site $E_T/L=\frac{T}{2}+2J\left[1-C_1\right]$. Since $E_{\rm in}=E_T$ the temperature satisfies
\be 
2J=\frac{T}{2}+2J\left[1-\frac{I_1(2J/T)}{I_0(2J/T)}\right]~.
\ee
Its solution is $x=4J/T=2.1312$ leading to the correlation length $\xi=1/\log(x)=1.321$ and correlators $C_R=x^{-R}$:
\be 
C_1=0.469,~C_2=0.220,~C_3=0.103,~C_4=0.049~,
\label{A:CR}
\ee
compare with Figure \ref{FigRelaxation}. Notice that the asymptotic thermal correlators after the sudden quench do not depend on $J\ll1$.

\end{document}